\documentclass[12pt]{iopart}
\newif\ifpdf
\ifx\pdfoutput\undefined
  \pdffalse
\else
  \pdfoutput=1
  \pdftrue
\fi
\ifpdf
   \usepackage[pdftex]{graphicx}
   \DeclareGraphicsExtensions{.pdf}
   \usepackage{thumbpdf}      
\else
   \usepackage[dvips]{graphicx}
   \DeclareGraphicsExtensions{.eps}
\fi
\usepackage{cite}
\usepackage{amssymb}
\usepackage{amscd}
\usepackage{subfigure}
\usepackage{setstack}
\usepackage{iopams}

\renewcommand{\text}[1]{{\mbox{#1}}}
\renewenvironment{pmatrix}{\left(\!\!\begin{array}{cc}}{\end{array}\!\!\right)}
\newcommand{\ri}{{ \rm i }}

\newcommand{\rR}{{ \rm R }}
\newcommand{\rL}{{ \rm L }}
\newcommand{\rA}{{ \rm A }}

\newcommand{\be}{\begin{equation}}
\newcommand{\ee}{\end{equation}}

\usepackage{color}
\definecolor{blau}{rgb}{0,0,1}
\definecolor{gruen}{rgb}{0,1,0}
\definecolor{rot}{rgb}{1,0,0}
\definecolor{magenta}{rgb}{1,0,1}

\begin{document}
\jl{1}
\title[Barrier transmission for the Nonlinear Schr\"odinger Equation]
{Barrier transmission for the Nonlinear Schr\"odinger Equation: Surprises of nonlinear transport}
\author{K Rapedius and H J Korsch}

\address{ FB Physik, TU Kaiserslautern, D-67653
Kaiserslautern, Germany}

\ead{korsch@physik.uni-kl.de}

\begin{abstract}
In this communication we report on a peculiar property of barrier transmission that systems
governed by the {\it nonlinear\/} Schr\"odinger equation share with the linear one:
For unit transmission the potential can be divided at an arbitrary point into two sub-potentials,
a left and a right one, which have exactly the same transmission.
This is a rare case of an exact property of a nonlinear wave function which will be
of interest, e.g., for studies of coherent transport of Bose-Einstein condensates
through mesoscopic waveguides.
\end{abstract}

\submitto{\JPA}

\vspace*{6mm}

\maketitle

Quantum mechanics is full of surprises, in some cases even in quite
elementary situations, that are seemingly well known from textbooks. A recent example is the following
remarkable observation for barrier penetration.

Let us consider the transmission of a
one-dimensional wavefunction through a potential $V(x)$ with $V(x)\rightarrow 0$ for
$x \rightarrow \pm \infty$. A well known example is a rectangular potential well
\be
V(x)=\left\{\begin{array}{cl}-V_0&,\ |x|\le a\\0&,\  |x|> a \end{array}\right.
\label{potwell}
\ee
($V_0>0$) with a transmission probability \cite{Merz70}
\be
|T|^2(E)=\left\{1+\frac{V_0^2}{4E(E+V_0)}\sin^2
\left(2a\sqrt{2m(E+V_0)\,}/\hbar\,\right)\right\}^{-1}
\ee
($m$ is the mass of the particle and $E$ the energy).
At certain ``resonance'' energies $E_{\rm res}$ the potential is $100\%$ transparent,
for example at $E_n=-V_0+(\hbar^2 \pi^2/8 m a^2)\, n^2>0$, $n=n_{\rm min},\ n_{\rm min}+1,\dots$,
for the rectangular well (\ref{potwell}). Such resonances have been studied for more general
potentials, as for example symmetrical or asymmetrical
double-barrier structures in connection with resonant-tunneling,
where they are denoted as unit resonances (see, e.g., \cite{Zhao95D} and references therein).

Chabanov and Zakhariev \cite{Chab06} discovered that, at such a resonant energy, the
potential scattering shows a surprising symmetry: Dividing $V(x)$
into two distinct parts at some arbitrary point $ x'$ defining a 'left' and a `right' potential
\be
    V_\rL(x)=\left\{\begin{array}{cl}
                     V(x)   &    x \le x' \\
                     0      &   x>x'\\
                     \end{array}
              \right.
\quad,\quad
    V_\rR(x)=\left\{\begin{array}{cl}
                     0     &    x < x' \\
                     V(x)  &   x \ge x'\\
                     \end{array}
              \right. \,,
              \label{pot-l-r}
\ee
the transmission probabilities $|T_\rL|^2$ for $V_\rL$ and $|T_\rR|^2$ for $V_\rR$
are equal at the resonance energy,
\be
  |T_\rL|^2(E_{\rm res})=|T_\rR|^2(E_{\rm res})\,,
\ee
despite the fact that the left and right potentials can be very different. This is shown
easily using the transfer matrix $\bf M$ connecting the amplitudes of the wavefunction
on the left hand side, $\psi(x)=A\exp(\ri kx)+B\exp(-\ri kx)$, with those on the right hand
side, $\psi(x)=C\exp(\ri kx)+D\exp(-\ri kx)$
(the limit $|x|\rightarrow \infty$ is understood for a potential not vanishing outside a finite range):
\be
\begin{pmatrix}C\\D\end{pmatrix}={\bf M}\begin{pmatrix}A\\B\end{pmatrix}=
\begin{pmatrix}\alpha&\beta\\\beta^*&\alpha^*\end{pmatrix}\begin{pmatrix}A\\B\end{pmatrix}
\ee
where the unit determinant $\det {\bf M}=|\alpha|^2-|\beta|^2=1$ guarantees that the reflection
probability $|R|^2=|\beta/\alpha|^2$ and the transmission probability 
$|T|^2=1/|\alpha|^2$ sum to unity (see, e.g., \cite{Merz70}). If the potential is cut at $x'$
according to (\ref{pot-l-r}), the total transfer matrix can be expressed as a product
of the transfer matrices ${\bf M}_\rR$ and ${\bf M}_\rL$ of the right and left potentials,
respectively: ${\bf M}={\bf M}_\rR{\bf M}_\rL$. For unit transmission we have $|\alpha|=1$ and
\be
\beta=\alpha_\rR\beta_\rL+\beta_\rR\alpha^*_\rL=0
\ee
which implies $\beta_\rR/\alpha_\rR=-\beta_\rL^*/\alpha_\rL$ and hence equal reflection
probabilities of $V_\rR$ and $V_\rL$.
This property, which can  also be extended to more general situations, offers
an intuitive approach to the design of potentials with desired transmission properties (see \cite{Chab06} and references therein for more details).

In this communication we report an even more puzzling fact, namely that these properties remain
valid for the nonlinear Schr\"odinger equation (NLSE)
\be
    \frac{\hbar^2}{2m} \psi''+(\mu-V)\psi-g|\psi|^2\psi=0\,.
    \label{1}
\ee
The nonlinearity destroys the superposition principle in linear quantum mechanics and we therefore
cannot base our proof on matrix techniques as above. 
The NLSE (\ref{1}), also known as Gross-Pitaevskii equation, attracted much interest in recent
years because it describes the dynamics of Bose-Einstein condensates in a mean field
approximation at low temperature (see, e.g., \cite{Legg01}). Note that the chemical potential $\mu$ takes over the
role of the energy $E$. We furthermore note that also the nonlinearity $g$ may be $x$-dependent.

Transport properties of cold atomic gases in designed mesoscopic waveguides are of
recent interest \cite{Paul07b}. Here in particular
barrier transmission of the nonlinear waves has been studied in a number of papers
\cite{Paul05,06nl_transport,Paul07b,07nlLorentz}
that assume an experimental setup in which matter waves from a large reservoir of condensed atoms at chemical potential $\mu$ are injected into a one-dimensional waveguide in which the condensate can propagate. In these articles it was shown that the results obtained from the stationary NLSE (\ref{1}) are in excellent agreement with numerical solutions of the time--dependent NLSE
\be
\fl
   \quad \ri \hbar \dot{\psi}(x,t)=-\frac{\hbar^2}{2m}\psi''(x,t)+V(x)\psi(x,t)
   +g|\psi(x,t)|^2\psi(x,t)+f_0\exp(-\ri \mu t/\hbar)\delta(x-x_0)\,
   \label{GPE_t}
\ee
where the source term $f_0 \exp(-\ri \mu t/\hbar)\delta(x-x_0)$ located at $x=x_0$ emits monochromatic matter waves at chemical potential $\mu$ and thus simulates the coupling to a reservoir.
The barrier potential $V(x)$ is assumed to be zero for $x \le x_0$.
As discussed in these articles,
the definition of a transmission probability is not unambiguous, because of the implicit
dependence on the scattering wave function $|\psi(x)|^2$.
Different prescriptions assuming either a fixed source strength $f_0$ or a fixed incoming current can be formulated which differ in a strongly nonlinear situation. No ambiguity appears, of course, if the nonlinearity vanishes in the
region far away from the scattering potential. Note that for unit transmission the differences
between fixed output or input conditions as well as the definition ambiguities disappear.
See \cite{Paul07b,07nlLorentz} for more details.

A fixed output boundary condition
\be
   \psi(x) \rightarrow C\exp(\ri kx),\,\,\, \psi'(x) \rightarrow \ri k C\exp(\ri kx) \text{ for } x \rightarrow + \infty \,
   \label{II}
\ee
with $k=\sqrt{2m(\mu-g|C|^2)}/\hbar$ determines the
wavefunction $\psi(x)$ and its derivative $\psi'(x)$ in the region $-\infty  \le x \le \infty$
uniquely \cite{Gred92}. This provides a recipe to define an effective incoming amplitude $A$ in the
upstream region and a transmission coefficient 
\be
|T|^2=(k/k_\rA)|C/A|^2\quad
\mbox{with}\quad k_\rA=\sqrt{2m(\mu-g|A|^2)}/\hbar\,.
\ee
For the fixed output problem, the transmission coefficient is a unique function of the
chemical potential $\mu$. If the problem is formulated as a fixed input problem, one
observes a bending over of the transmission spectra and an interaction induced
bistability for strong nonlinearity
where the transmission coefficient is no longer a unique function of $\mu$.
In addition, it was found that the points of unit transmission, the resonances $\mu_{\rm res}$,
survive in the nonlinear case $g\ne 0$, of course shifted to different positions depending on $g$.
In the following examples, we will use transmission coefficients derived from the fixed
output condition (\ref{II}).

If the potential $V(x)$ is $100\%$ transparent at the chemical 
potential $\mu=\mu_{\rm res}$, the solution in the far upstream region is given by
\be
   \psi(x) \rightarrow C\exp(\ri kx +\ri  \varphi),\,\,\, \psi'(x) \rightarrow \ri k C\exp(\ri kx +\ri  \varphi) \text{ for } x \rightarrow - \infty \, .
   \label{IVa}
\ee
where $\varphi$ is a real-valued phase.

Now we define a second fixed output problem given by the NLSE
\be
    \frac{\hbar^2}{2m} \chi''+(\mu-V)\chi-g|\chi|^2\chi=0
    \label{2}
\ee
with the initial conditions
\be
   \chi(x) \rightarrow C^*\exp(-\ri kx-\ri \varphi),\,\,\, \chi'(x) \rightarrow -\ri k C^*\exp(-\ri kx -\ri \varphi) 
   \label{IVb}
\ee
for $x \rightarrow - \infty$, which corresponds to scattering through $V(x)$ from the opposite side of the well.
Noting that, for real values of $\mu$, the complex conjugate of a solution also solves the NLSE, we find immediately
that the solution of (\ref{2}) and (\ref{IVb}) is given by
\be
   \chi(x)=\psi^*(x) ,\,\,\, \chi'(x)=\psi'^*(x),
   \label{V}
\ee
and in particular
\be
   \chi(x)=C^* \exp(-\ri k x),\,\,\, \chi'(x)=-\ri k C^*\exp(-\ri k x) \text{ for } x \rightarrow + \infty \, .
\ee
This means that the barrier potential $V(x)$ is also transparent for waves with chemical potential $\mu$ coming
in from the opposite side of the barrier even if the potential is asymmetrical. Note that this is not satisfied
automatically, because the equality of the left-to-right and right-to-left transmission probabilities
valid in the linear case is not guaranteed for the NLSE, at least to the knowledge of the authors.

Now we divide the barrier potential $V(x)$ as given in equation (\ref{pot-l-r}) and
obtain two new fixed output problems:
The first one is determined by the NLSE
\be
    \frac{\hbar^2}{2m} \psi_\rR''+(\mu-V_\rR)\psi_\rR-g|\psi_\rR|^2\psi_\rR=0
    \label{3}
\ee
with the initial conditions
\be
   \psi_\rR(x) \rightarrow C\exp(\ri kx),\,\,\, \psi_\rR'(x) \rightarrow \ri k C\exp(\ri kx) \text{ for } x \rightarrow + \infty \, ,
   \label{4}
\ee
which corresponds to tunneling through the potential $V_\rR(x)$ from the left to the right hand side of the barrier.
The second one is determined by the NLSE
\be
    \frac{\hbar^2}{2m} \chi_\rL''+(\mu-V_\rL)\chi_\rL-g|\chi_\rL|^2\chi_\rL=0
    \label{5}
\ee
with the initial conditions
\be
   \chi_\rL(x) \rightarrow C^*\exp(-\ri kx-\ri \varphi),\,\,\, \chi_\rL' \rightarrow -\ri k C^*\exp(-\ri kx -\ri \varphi)
   \label{IVc}
\ee
for $x \rightarrow - \infty$, which corresponds to tunneling through the potential $V_\rL(x)$ from the right to the left hand side of the barrier.

From equation (\ref{V}) we obtain at the cutting position $x'$
\be
   \chi_\rL(x')=\psi_\rR^*(x'), \, \, \, \chi_\rL'(x')=\psi_\rR'^*(x')\,.
   \label{x'}
\ee
Since the wavefunctions in the respective potential--free regions are uniquely determined by the values of the wavefunction and its derivative at $x'$ it follows together with the initial conditions (\ref{4}) and (\ref{IVc}) that the asymptotic wavefunction $\chi_\rL(x)$ for  $x\rightarrow \pm \infty$ is equal to the asymptotic wavefunction $\psi_\rR^*(x)$ for $x \rightarrow \mp \infty$ up to an irrelevant phase factor $\exp(\ri \varphi)$. Thus the respective transmission coefficients coincide no matter which particular definition of the transmission coefficient is used.

For illustrative purposes we explicitly demonstrate in the following the equality of the two transmission coefficients for the case that a fixed source strength $f_0$ is assumed in equation (\ref{GPE_t}). In section II.A of \cite{07nlLorentz} it was shown that the source strength is connected with the effective incoming amplitude $A_\rR$ via $f_0=\ri \frac{\hbar^2}{m}k_\rR A_\rR$ with $k_\rR=\sqrt{2m(\mu-g|A_\rR|^2)}/\hbar$. By considering the stationary solutions of equation (\ref{GPE_t}) it was further shown that $A_\rR$ is determined by the wavefunction and its derivative at the position $x_0$ of the source. Choosing $x_0=x'$ we obtain from equation (12) in \cite{07nlLorentz}
\be
   2 \ri k_\rR A_\rR=\psi_\rR'(x') + \ri k' \psi_\rR(x')
   \label{2ikA}
\ee
with $k'=\sqrt{2m(\mu-g|\psi_\rR(x')|^2)}/\hbar=\sqrt{2m(\mu-g|\chi_\rL(x')|^2)}/\hbar$.
Since $\chi_\rL(x)$ corresponds to scattering from the opposite direction the respective equation for the effective incoming amplitude $A_\rL$ differs by a sign.
With the definition $k_\rL=\sqrt{2m(\mu-g|A_\rL|^2)}/\hbar$ we thus arrive at
\be
   2 \ri k_\rL A_\rL=\chi_\rL'(x') - \ri k' \chi_\rL(x')=\big(\psi_\rR'(x') + \ri k' \psi_\rR(x')\big)^*=\big(2 \ri k_\rR
   A_\rR \big)^*
\ee
using (\ref{x'}) and (\ref{2ikA}).
Consequently, we obtain $|A_\rL|^2=|A_\rR|^2$ and $k_\rL=k_\rR$ so that the respective transmission coefficients
\be
   |T_\rL|^2=\frac{k|C|^2}{k_\rL |A_\rL|^2}=\frac{k|C|^2}{k_\rR |A_\rR|^2}=|T_\rR|^2
\ee
coincide. More precisely, the left-to-right transmission of $V_\rL$ is equal to the
right-to-left transmission of $V_\rR$ (compare the remark above). 

\begin{figure}[htb]
\centering
\includegraphics[width=7cm,  angle=0]{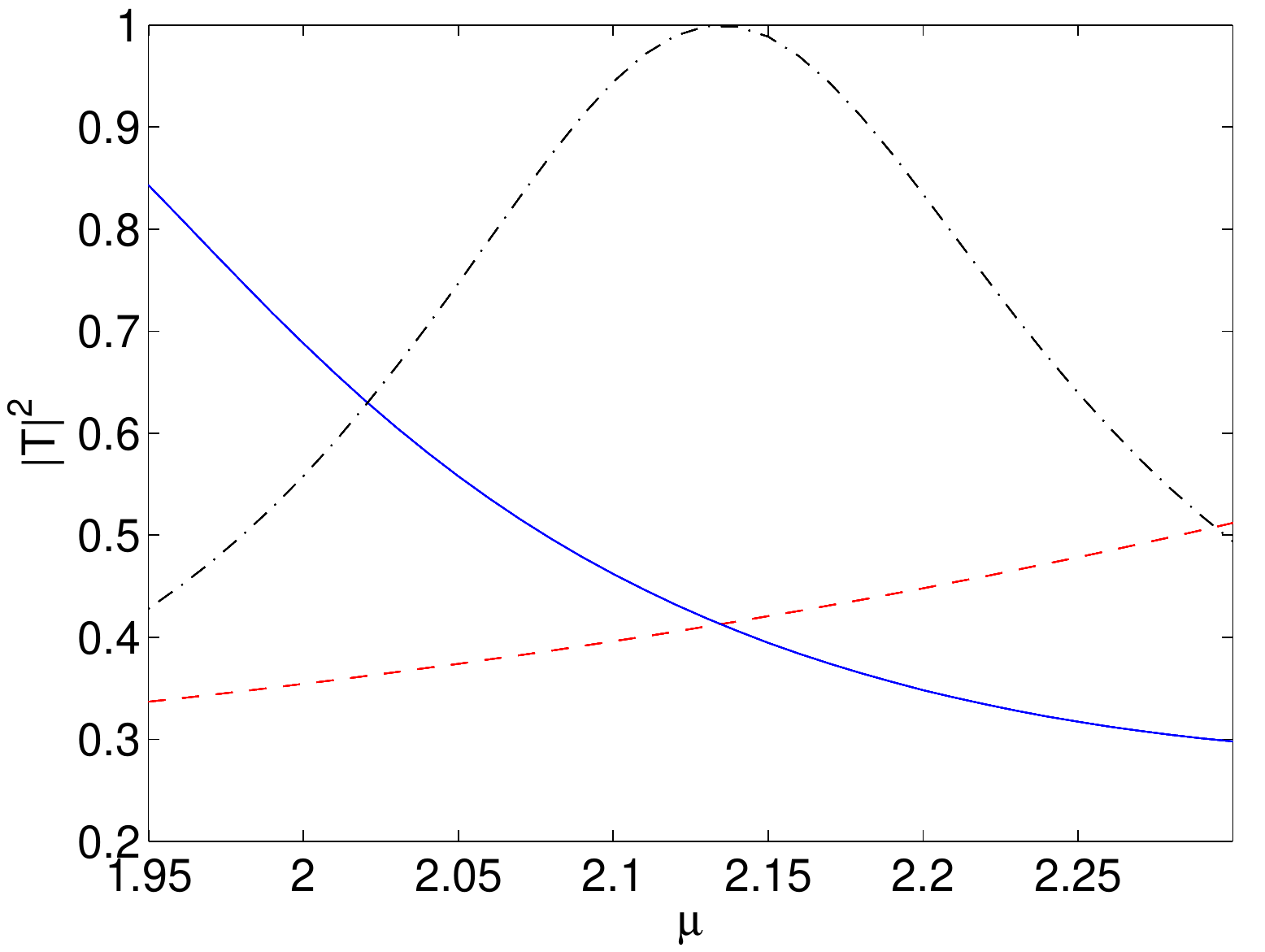}
\caption{\label{fig-square_out} Transmission coefficient (-.-.) of the NLSE for the rectangular
potential well (\ref{potwell}) (parameters
$m=\hbar=1$, depth $V_0=50$, width $2a=40$ and attractive nonlinearity $g=-1$).
Also shown are the transmission probabilities for the left (width $10$, \textcolor{rot}{$---$}) and right
potential (width $30$,  \textcolor{blau}{$-$}) obtained from dividing $V(x)$ into two parts. These two
probabilities are equal at the resonance at $\mu_{\rm res}$ where the potential $V(x)$ is
transparent.}
\end{figure}

As an illustration, we present results based on a numerical solution of the NLSE
for two cases.
Figure \ref{fig-square_out} shows the transmission probability
for a rectangular well (\ref{potwell}) with $m=\hbar=1$, $V_0=50$, $a=20$ and
an attractive nonlinearity $g=-1$ acting along the whole $x$-axis.
The outgoing boundary conditions are chosen as $C=1$.
We observe unit transmission, $|T|^2=1$, at the resonance $\mu_{\rm res}\approx 2.135$.

We now divide the rectangular well with width $2a=40$ at the value $x'=-10$ into two separate parts and
calculate the transmission coefficient
of each of these rectangular wells $V_\rL$ (width 10) and $V_\rR$ (width 30) separately. Their 
transmission curves displayed in figure \ref{fig-square_out}  cross at the chemical potential $\mu_{\rm res}$,
exactly as predicted.

\begin{figure}[htb]
\centering
\includegraphics[width=7cm,  angle=0]{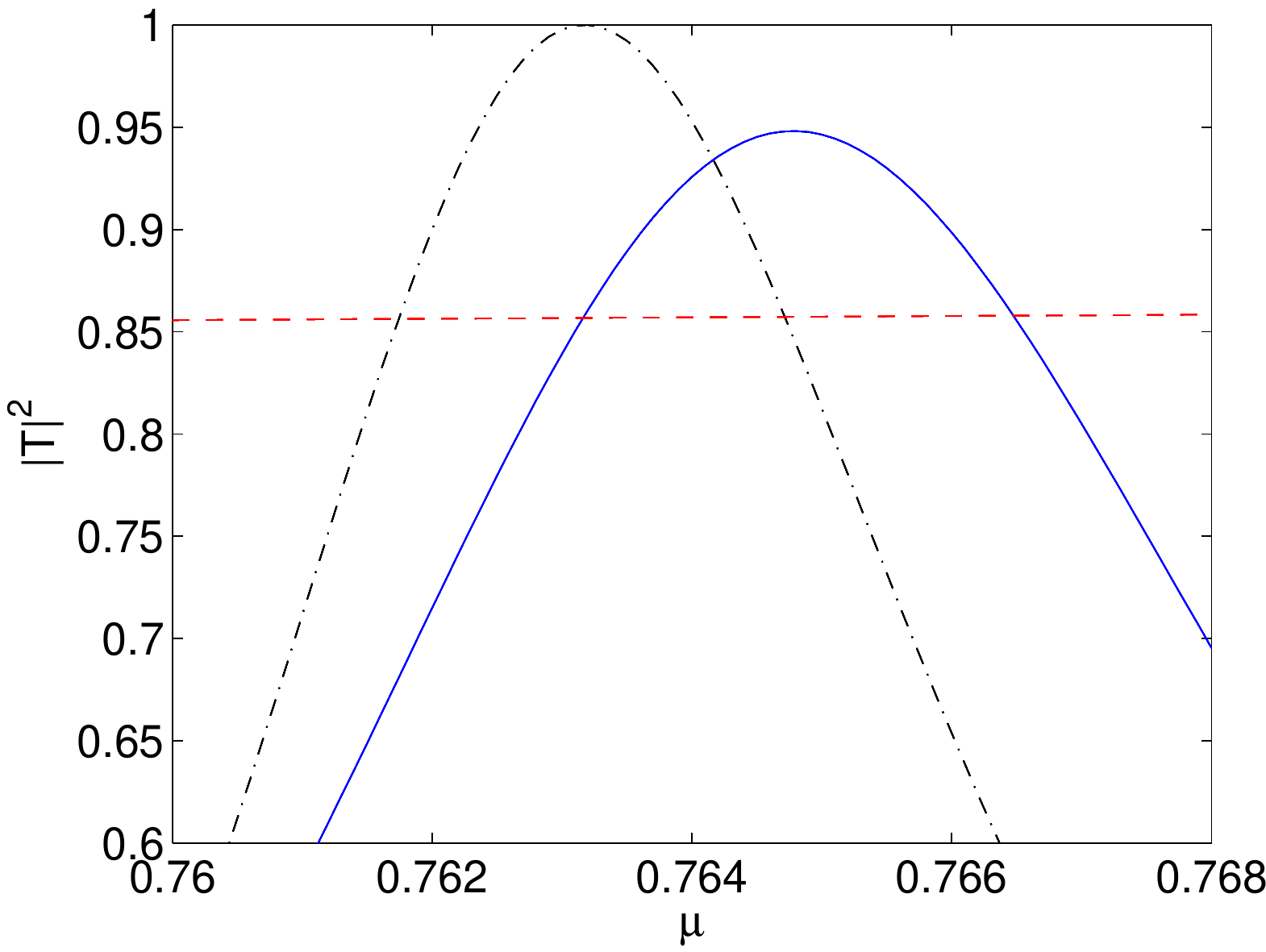}
\caption{\label{fig-DGauss_out} Same as figure \ref{fig-square_out}, however for the double
barrier potential (\ref{DGauss}).}
\end{figure}

As a second example we choose the double-Gaussian barrier 
\be
   V(x)=V_0\left[\exp(-(x+b)^2/\alpha^2)+\exp(-(x-b)^2/\alpha^2)\right]
   \label{DGauss}
\ee
with parameters $V_0=1$, $b=7.35$, $\alpha=b/5$ and a nonlinearity of $g=0.005$ studied
in \cite{Paul05,07nlLorentz}, where the apparently small value of $g$ nevertheless causes
a strong deviation of the nonlinear resonance curve from the linear one. In figure \ref{fig-DGauss_out} we observe full transparency at $\mu \approx 0.7632$, i.e.~below the potential maxima.
In addition the figure shows the transmission curves for the left and right potentials, where the potential is divided by $x=-8.0$. Again the left and right transmission coefficients are equal at the
point of transparency.

In conclusion, we have shown that the interesting property of linear quantum barrier
penetration at unit transmission, the equality of transmission through the arbitrarily divided sub-potentials,
can be carried over to the nonlinear case which is much less understood.
In addition
of being of interest in its own right, this remarkable fact may provide
an intuitive way to treat transport in nonlinear quantum systems, as for example
the Bose-Einstein condensates in the mean-field approximation.

\ack
Support from the Deutsche
Forschungsgemeinschaft via the Graduiertenkolleg ''Nichtlineare Optik
und Ultrakurzzeitphysik'' is gratefully acknowledged.

\section*{References}

\end{document}